# Understanding Phonon Scattering by Nano-Precipitates in Potassium-Doped Lead Chalcogenides

Zhao Wang[a,b], Xiaolong Yang[b], Dan Feng[b], Haijun Wu[c], Jesus Carrete[d], Li-Dong Zhao[e], Chao Li[f], Shaodong Cheng[f], Biaolin Peng[a], Guang Yang[f]* and Jiaqing He[g]*

[a] Guangxi Key Laboratory for Relativistic Astrophysics, Department of Physics, Guangxi University, Nanning 530004, P. R. China.

[b] Frontier Institute of Science and Technology, and State Key Laboratory for Mechanical Behavior of Materials, Xi'an Jiaotong University, 710054, Xi'an, P. R. China.

[c] Department of Materials Science and Engineering, National University of Singapore, 117546, Singapore.

[d] LITEN, CEA-Grenoble, 17 rue des Martyrs, 38054 Grenoble Cedex 9, France.

[e] School of Materials Science and Engineering, Beihang University, Beijing 100191, P. R. China.

[f] International Center for Dielectric Research, Xi'an Jiaotong University, Xi'an 710049, P. R. China.

[g] Department of Physics, South University of Science and Technology of China, Shenzhen 518055, P. R. China.

ABSTRACT: We present a comprehensive experimental and theoretical study of phonon scattering by nanoprecipitates in potassium-doped PbTe, PbSe and PbS. We highlight the role of the precipitate size distribution measured by microscopy, whose tuning allows for thermal conductivities lower than the limit achievable with a single size. The correlation between the size distribution and the contributions to thermal conductivity from phonons in different frequency ranges provides a physical basis to the experimentally measured thermal conductivities, and a criterion to estimate the lowest achievable thermal conductivity. The results have clear implications for efficiency enhancements in nanostructured bulk thermoelectrics.

# Introduction

Thermoelectric energy conversion shows a number of clear advantages over more conventional thermodynamic cycles. Scalability to reduced sizes, lack of moving parts and quiet operation figure chiefly among them.[1-3] Its applicability has, nonetheless, been hindered so far by their low efficiencies, a problem that is only slowly being allayed. The efficiency of a thermoelectric material is correlated to its dimensionless figure of merit $ZT = (S^2\sigma/\kappa)T$, where $S$, $\sigma$, $\kappa$ and $T$ are the Seebeck coefficient, electrical conductivity, thermal conductivity, and absolute temperature, respectively. Among the possible strategies for improving $ZT$, one of the most fruitful so far has been based on reducing $\kappa$. Numerous strategies to include multiscale phonon scattering sources in bulk thermoelectrics have been proposed, including substitution doping, precipitation, and nanograins.[4-8] Among these, nanometer-sized precipitates (so-called nanoprecipitates) are particularly attractive due to their potential simplicity of manufacturing by means of ball-milling, continuous gas-phase, hot-pressing, or spark plasma sintering.[9] The effect of nanoprecipitates on thermal conduction has thus drawn much attention in the search for efficient nanostructured thermoelectric materials.[10-12]

Lead chalcogenides are a well-known family of high-efficiency thermoelectrics based on Earth-abundant elements.[13-15] A classical approach to improve their ZT involves doping them with an appropriate amount of impurities so as to optimize the electronic carrier concentration.[8,16-19] Some remarkable peak ZTs have been reported for 1%Al-doped PbSe,[20] 2%Na-doped PbTe-PbS[8] and 1.25%K-doped PbSe:[16] about 1.3, 1.5 and 1.25 at about 850, 750 and 850 K, respectively. Such enhancements in ZT have been found to come not only from the optimization of charge carriers, but also from a strong reduction in thermal conductivity caused by doping-induced nanostructures acting as effective phonon scattering centers.[21] Although those results are revealing, most previous works resorted to either classical transport property analyses[8,16,20] or simple nanostructure measurements.[5, 22] The correlation between the enhancement of ZT and the presence of nanostructures related to the doping process was therefore not well understood at the time, mainly due to lack of appropriate theoretical modeling of transport properties based directly on experimental nanostructure measurements. In particular, most modeling efforts rely on the assumption of a single average nanoparticle size,[22-28] The possible underestimation of phonon scattering by nanoprecipitates stemming from that hypothesis deserves more attention, given the fact that scattering centers with sizes distributed across a range can scatter phonons more efficiently than monodisperse ones.[26]

Only very recently has it become possible to precisely measure the size distribution of nanoprecipitates in experiments,[23,25-28] with the above motivations it is crucial to bridge this with the behavior of thermoelectric transport coefficients via theoretical tools in view to enhancing thermoelectric performance. To this end, we synthesized potassium-doped PbTe, PbSe and PbS composite samples with high-density nanoprecipitates observed by transmission electron microscopy (TEM). Those samples were prepared to have intentionally large grain sizes so as to avoid a dominating effect of grain boundary scattering and thus simplify the analysis. We measured the thermoelectric transport coefficients, and here we provide TEM sample micro-structure distribution statistics, and try to correlate them by a new theoretical approach beyond the previously proposed single-size model.[24]

## Experimental Methods

### Sample synthesis

Three samples of $Pb_{0.98}K_{0.02}Q$ (Q = Te, Se or S) were prepared by a melting reaction by mixing elemental Pb, Q and K within carbon-coated fused silica tubes. The reaction tubes were evacuated to a base pressure of ~$10^{-4}$ torr, flame-sealed, slowly heated to 723 K over 12 h and then to 1423 K over 7 hours, soaked at this temperature for 6 hours and subsequently air-quenched to room temperature.

Reagent chemicals were used as obtained: Pb wire (99.99%, American Elements, US), Te shot (99.999%, 5 N Plus, Canada), Se shot (99.999%, 5 N Plus, Canada), S chunk (99.999%, Inc., Canada) and K chunk (99.999%, Aldrich, US).

**Measurement of thermoelectric transport coefficients**

The obtained ingots were cut into bars of 15×3×2 mm, which were used for measurements of the Seebeck coefficient $S$ and electrical conductivity $\sigma$ using an Ulvac Riko ZEM-3 instrument under a helium atmosphere from about 300 to 923 K. Electrical transport coefficients measured from different slices cut from the same pellets were found to be close, attesting to a good sample homogeneity. The uncertainty of $S$ and $\sigma$ was estimated to be within about 5%.

The highly dense ingots were cut and polished into coins with Ø ~ 8 mm and 1~2 mm thicknesses for thermal diffusivity ($D$) measurements using laser flash diffusivity in a Netzsch LFA457 instrument. The samples were coated with a thin layer of graphite to minimize errors in the emissivity of the material. The measured data were analyzed using a Cowan model with pulse correction, and heating and cooling cycles were found to yield a repeatable diffusivity for each given sample. Thermal diffusivity values obtained for different slices from the same pellet were close. The thermal conductivity $\kappa$ is calculated with the relation $\kappa=DC_p\rho_m$, where $\rho_m$ is the mass density and $C_p$ is the specific heat capacity. $C_p$ was indirectly derived using a standard sample (Pyroceram 9606) at temperatures ranging from 300 to 923K (results shown in Supporting Information). Our $C_p$ results are in good agreement with previously reported values.[29] The sample density $\rho$ was determined by gas pycnometry (Micromeritics AccuPyc1340) measurements. The uncertainty of thermal conductivity is estimated to be within 8% by considering a combination of uncertainties for $D$, $C_p$ and $\rho_m$.

**Microscopy**

**Transmission Electron microscopy (TEM)**

TEM nanostructure measurements were carried out in a JEOL 2100F and a JEM-ARM200 microscopes. The bulk TEM specimens were prepared by standard methods including cutting, grinding, dimpling, polishing and Ar-ion milling in a liquid-nitrogen-cooled environment. Energy-dispersion X-ray spectroscopy analyses were performed to determine the possible composition of nanoprecipitates. We note that we cannot expect nanoparticles to be thermodynamically stable at high temperatures from a rigorous point of view. It is known that the usual high-energy ion beam will cause beam damage to the samples, especially in the area close to the edge. In order to eliminate the beam damage artifacts from our interpretation, we milled the sample at the temperature of liquid $N_2$, we polished the samples very thin (below 30 microns) so that very short ion milling times were needed (about 30 min) and milled the samples at a low angle (6 degrees) and low energy (1 kV) as a final step to further reduce the extent of ion milling damage.

**Optical microscopy**

The PbTe sample was carefully grinded and polished free of surface scratches, then progressively heated to about 573K in a sintering furnace. The temperature was maintained for 2 minutes before cooling to room temperature. The grain size distribution was measured in a Nikon-Lv150n optical microscope.

**Electron Backscattered Diffraction (EBSD)**

The EBSD analysis was performed using an FEI Quanta 250F environmental scanning electron microscope (SEM) equipped with an EDAX Hikari camera for capturing the EBSD patterns. For EBSD pattern acquisition, the microscope was operated at a 20 kV accelerating voltage and 9.5 mm working distance with the sample tilted to 70 degrees (the maximum allowed tilt for this microscope) using the EDAX-TEAM EBSD data acquisition and data analysis software from EDAX. The EBSD system was carefully calibrated at the given working distance before the performing the experiments.

## Theoretical analyses

Under the relaxation-time approximation (RTA), [30-31] the lattice thermal conductivity of an isotropic material can be expressed as

$$\kappa = \frac{1}{3}\sum_i \int \frac{d^3q}{8\pi^3} v_{iq}^2 \tau_{iq} c_{iq}, \qquad (1)$$

where the sum runs over all phonon bands (i), the integral extends over the first Brillouin zone, $v_{iq}$ is the group velocity of a given phonon mode, $\tau_{iq}$ is its relaxation time, and $c_{iq}$ its contribution to the volumetric heat capacity. We provide detailed information about the phonon spectrum calculation and a detailed comparison to previous theoretical calculations as well as with experimental data as a part of the Supporting Information. The total relaxation time can be approximated by a Matthiessen-type sum of the Umklapp ($\tau_u$), impurity ($\tau_i$), boundary ($\tau_b$), and nanoparticle ($\tau_{np}$) contributions,

$$\tau^{-1} = \tau_u^{-1} + \tau_i^{-1} + \tau_b^{-1} + \tau_{np}^{-1}. \qquad (2)$$

The Umklapp contribution is described by [21, 24]

$$\tau_u^{-1} = \varphi \omega^2 \frac{T}{T_D} e^{\frac{-T_D}{3T}}. \qquad (3)$$

Here the parameter $\varphi$ is obtained by a fit to our experimental lattice thermal conductivity data. We obtained the Debye temperature $T_D$ from the second moment of the phonon frequency $\omega$ distribution.[32] The scattering rate contributed by impurity scattering can be written as [33]

$$\tau_i^{-1} = \delta^3 c_i \left(\frac{m_i - m_{ave}}{m_{ave}}\right)^2 \frac{\omega^4}{4\pi v^3}, \tag{4}$$

where $\delta^3$ is the volume around one atom in the lattice, $c_i$ is the point defect concentration, $m_i$ is the mass of a point defect and $m_{ave}$ is the average mass of the crystal. The boundary scattering term is estimated as $v_{iq}$ divided by the average grain size $l$ times a form factor $\xi$

$$\tau_b^{-1} = \xi \frac{v_{iq}}{l}. \tag{5}$$

Here $\xi$ is fitted to our experimental lattice thermal conductivity data. The precipitate scattering term $\tau_{np}$ is computed by incorporating our measured precipitate size distribution into an expression for the scattering cross-section $\theta$ as an interpolation between a $\propto \omega^4$ Rayleigh-like long-wavelength regime $\theta_l$ [34]

$$\theta_l = \frac{1}{9}\pi D^2 [(\Delta\rho/\rho)^2 + 12(\Delta K/K)^2]\left(\frac{\omega D}{2v_g}\right)^4, \tag{6}$$

and a frequency-independent short-wavelength geometric limit $\theta_s$,

$$\theta_s = \frac{\pi D^2}{2}, \tag{7}$$

where $D$ is the particle diameter, $\rho$ is the mass density of the matrix material and $\Delta\rho$ is the mass density difference between the precipitate and matrix materials, $K$ is the force constant of the matrix material and $\Delta K$ is the force-constants difference between the precipitate and matrix, where they are replaced by elastic constants.[35]

In previous analyses considering only a single average precipitate size,[23-27] $\theta$ was simply taken as a Matthiessen-type combination as follows,

$$\theta^{-1} = \theta_s^{-1} + \theta_l^{-1}. \tag{8}$$

However, recent experiments have found that the size of nanoprecipitates is not tightly concentrated, but distributed in a certain length scale range in general from 0.5 to 20 nm.[22-23, 25-28] Thus, Eq.(8) above will fail to include the effect of hierarchically distributed nanostructures that scatter phonons more efficiently than those in a single length scale.[26, 36] To improve this situation, we take instead

$$\theta = \int_0^\infty (\theta_s^{-1} + \theta_l^{-1})^{-1} f(D) dD, \qquad (9)$$

where $f(D)$ is a probability density function describing the distribution of precipitate size. Knowledge of $\theta$ is enough to determine the precipitate contribution to scattering rate,

$$\tau_{np}^{-1} = v_g \theta N, \qquad (10)$$

where $N$ is the precipitate number density and $v_g$ is the phonon group velocity.

# Results and discussion

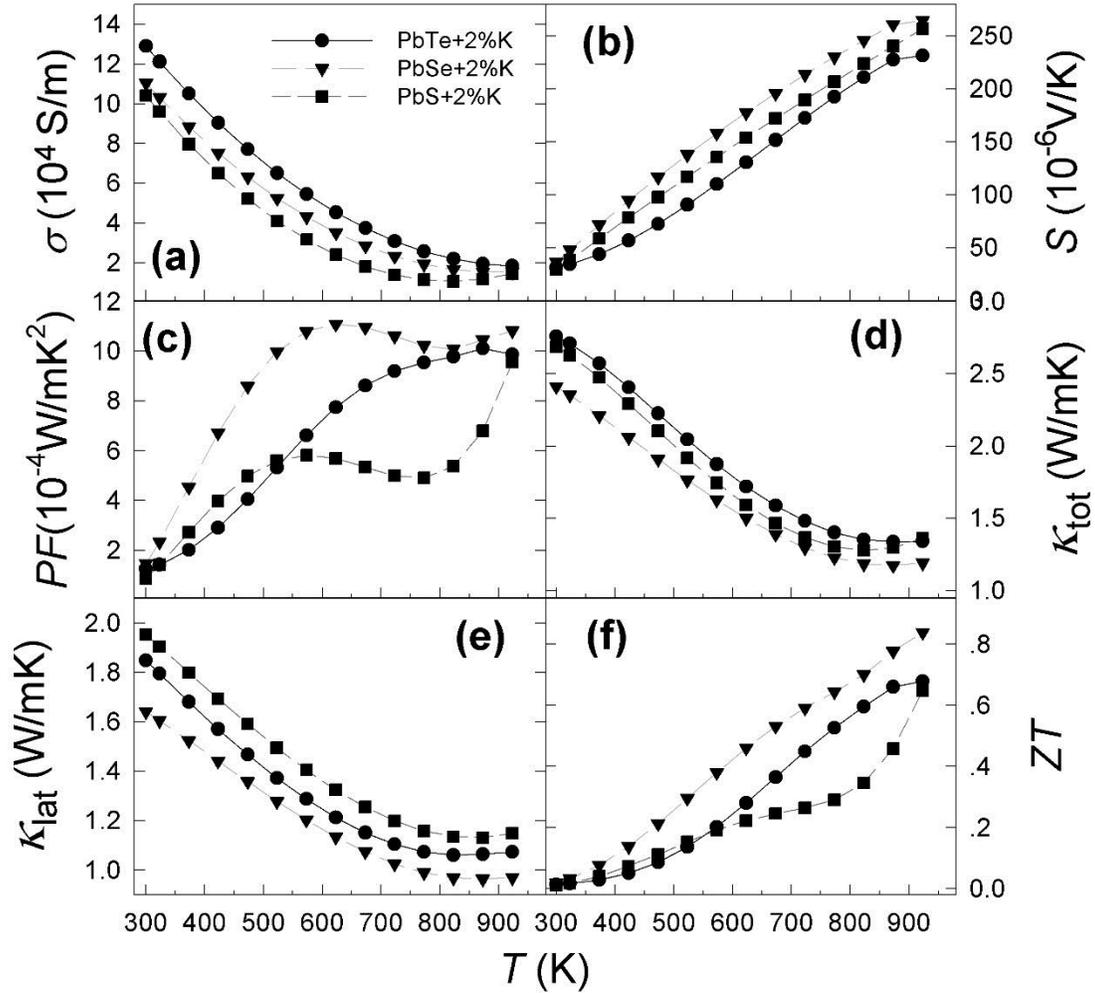

**Figure 1.** Thermoelectric transport coefficients versus temperature $T$ for three different samples: 2mol%-K-doped PbTe (PbTe+2%K), 2mol%-K-doped PbSe (PbSe+2%K) and 2mol%-K-doped PbS (PbS+2%K). (a) Total thermal conductivity $\kappa_{tot}$. (b) Power factor $PF=S^2\sigma$. (c) $ZT= PF*T/\kappa_{tot}$. (d) Electrical conductivity $\sigma$. (e) Seebeck coefficient $S$. (f) Lattice thermal conductivity $\kappa_l$.

We plot the thermoelectric transport coefficients of the K-doped PbTe, PbSe and PbS samples as a function of temperature in Figure 1. It can be seen that the total thermal conductivity decreases from 2.8 to 1.5 Wm$^{-1}$K$^{-1}$ with increasing temperature, with slight differences between samples (Figure 1d). The power factor increases to maximum values of 10.2, 11.3 and 9.4 10$^{-4}$W/mK$^2$ at 850, 650 and 900 K for the PbTe+2%K, PbSe+2%K and PbS+2%K samples, respectively (Figure 1c). Those result in peak *ZT* values of 0.65, 0.85 and 0.63 at 900K for the PbTe+2%K, PbSe+2%K and PbS+2%K samples, respectively (Figure 1f). In the literature, Zhang *et al.* reported peak *ZT* values of ~1.2 for $K_{0.0125}Pb_{0.9875}Se$ (at ~900K) and $K_{0.015}Pb_{0.985}Te$ (at ~700K),[16] and Girard *et al.* reported a peak *ZT* of ~1.1 for 2%Na-doped PbTe at ~800K.[8] Comparing our data with those as well as with results reported for pristine lead chalcogenides,[37] we see that our lower *ZT* is mainly owing to the lower electrical conductivities (Figure 1a) and Seebeck coefficients (Figure 1b) of our samples. These clearly indicate that two percent is not the optimal potassium doping portion. However, our samples exhibit lower lattice thermal conductivities than those reported for sodium-doped lead chalcogenides as shown below.[8, 25] We note that, in a simplistic analysis in the framework of the rigid-band approximation, doping concentration can be simply assumed to act as a knob to tune the position of the Fermi level. This approximation is valid in some cases, but recent experiments have shown that nanostructured bulk materials challenge the expectations of this simplified approximation.[20, 38] Specifically, under strong doping the dopant starts to strongly influence nanostructural features such as precipitate size and distribution.

To calculated the lattice contribution $\kappa_l$ to the total thermal conductivity, we estimate the Lorenz number *L* from a reduced chemical potential $\eta$ assuming a single, parabolic band and a dominant role of acoustic phonons in electronic scattering,[39]

$$L = \left(\frac{k_B}{e}\right)^2 \frac{3F_0(\eta)F_2(\eta) - 4F_1(\eta)^2}{F_0(\eta)^2}, \qquad (11)$$

where $k_B$ is the Boltzmann constant, *e* is the elementary electrical charge. $\eta$ is obtained from a fit to the experimental Seebeck coefficients,

$$S = \frac{k_B}{e}\left[\frac{2F_1(\eta)}{F_0(\eta)} - \eta\right], \qquad (12)$$

with $F(\eta)$ an n-order Fermi integral,

$$F_n(\eta) = \int_0^{+\infty} x^n (e^{x-\eta} + 1)^{-1} dx. \qquad (13)$$

$L$ is used to calculate $\kappa_l$ using the Wiedemann–Franz relation,

$$\kappa_l = \kappa_{tot} - L\sigma T. \qquad (14)$$

The lattice thermal conductivity values are plotted in Figure 1e. We see that $\kappa_l$ decreases with increasing temperature. The data range is roughly consistent with those reported for lead chalcogenides doped with K,[25] Na,[5, 8] Al,[25] and other elements.[40-42] We also see a slight increase of $\kappa_l$ above 850K. This implies possible bipolar effects that may correspond to the plateaus of Seebeck coefficient at high temperatures.[4]

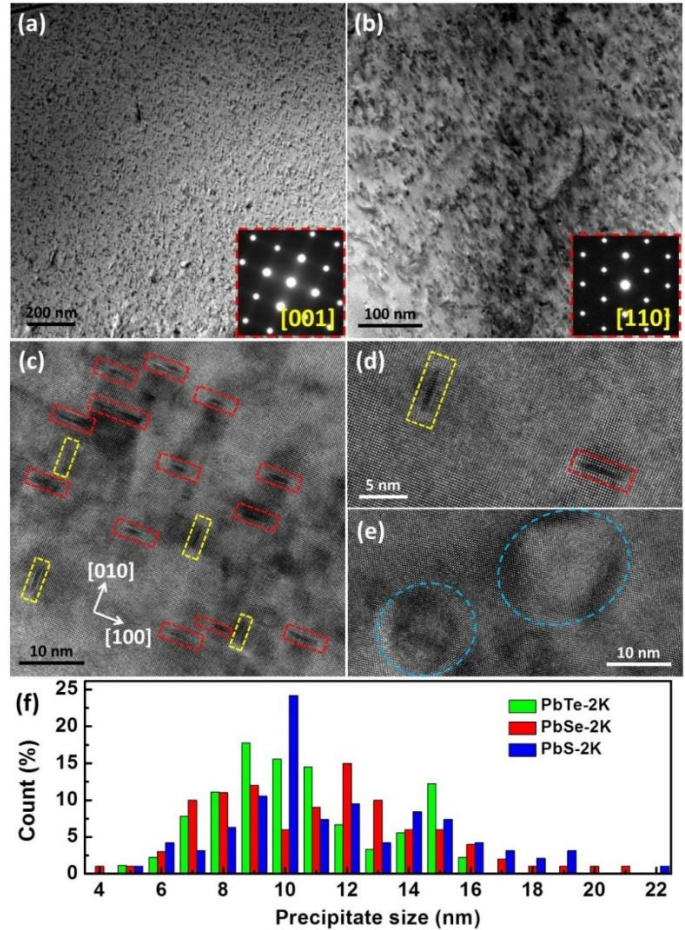

**Figure 2.** Microstructures of nanoprecipitates. (a,b) TEM images show high densities of nanoprecipitates along the [001] and [110] zone axes for PbTe+2%K and PbS+2%K samples, respectively; the insets are electron diffraction patterns; (c) HRTEM image along the [001] zone axis for PbTe+2%K depicting many perpendicular or parallel platelet-like precipitates along [100] directions; (d) Enlarged lattice image from (c) showing two perpendicular platelet-like precipitates; (e) HRTEM image along [110] zone axis for PbS+2%K reflects elliptical/round precipitates; (f) Size distribution histogram of nanoprecipitates in PbTe+2%K, PbSe+2%K, PbS+2%K samples.

The topology, density and distribution of nanoscale precipitates are shown in Figure 2. We see that nanoprecipitates are widely and densely distributed across the samples (Figure 2a and 2b from different observation directions). Figure 2c and *d* show the platelet-like topology of nanoprecipitates in our samples, which is in agreement with previous experimental observations on lead chalcogenide thermoelectrics,[4, 7, 25-26]: When observed from the [100] direction (Figure 2e), precipitates appear as two perpendicular dark lines with a width of one or two atom layers, whereas they present spherical or ellipsoidal cross sections when observed along non-[100] directions. This anisotropic morphology indicates a preferential direction of growth. Figure 2*f* shows the precipitate size distributions for PbQ+2%K using the same TEM observation volume. We see that the precipitate size and density are similar in all PbQ+2%K samples (see Supporting Information), with an average particle size ranging from 4 to 22 nm. The estimated (within 5% of error) distribution density of nanoprecipitates ranges from 0.8 to $1.2 \times 10^{12}$ cm$^{-2}$.

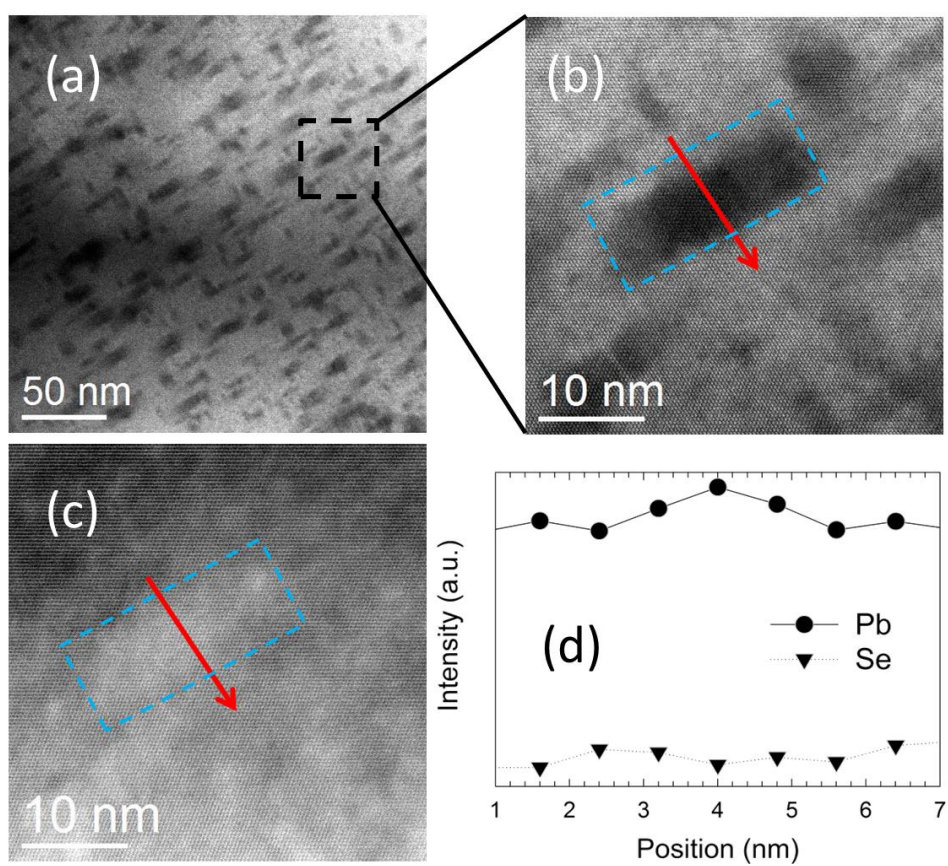

**Figure 3.** (a) Annular bright field (ABF) image of the PbSe+2%K sample. (b) High-magnification ABF image of the area selected in (a). The red arrow indicates the location and direction of the EDS line scan. (c) High-angle annular dark field (HAADF) image of the same area as in (b), with different contrast to make the precipitate appear bright. (d) EDS line scan intensity profile of the elements along the line indicated in (b) and (c).

To check the possible composition of the nanoprecipitates, high-magnification atomic imaging and energy dispersive spectroscopy (EDS) analyses are performed. Figure 3a shows the scanning transmission electron micrograph (STEM) annular bright field (ABF) image of PbSe-2k. The ABF image in Figure 3a depicts a high density of weak-contrast nanoscale rectangular and dark line-shaped precipitates. All precipitates are organized as observed in TEM (Figure 3). The ABF image (Figure 3b) and high-angle annular dark field (HAADF) image (Figure 3c) from the same area show precipitates (darker in the bright field image, brighter in the dark field image) with more high-Z elements. The EDS line scan along the red line in Figures 3b and 3c qualitatively indicates an apparent increase in the Pb signal from the precipitates compared with the matrix region (Figure 3d). This suggests that the precipitates are rich in Pb. We cannot deduce the exact chemical composition of the precipitates due to the existence of unavoidable matrix signal, i.e. the EDS signals were from the whole illuminated volume, not only including the precipitates but also the matrix surrounding and beneath them. The Se and K signals are also observed in the EDS results from the precipitates possibly due to the large x-ray interaction volume where signals from both precipitates and matrix contribute to the final spectra.

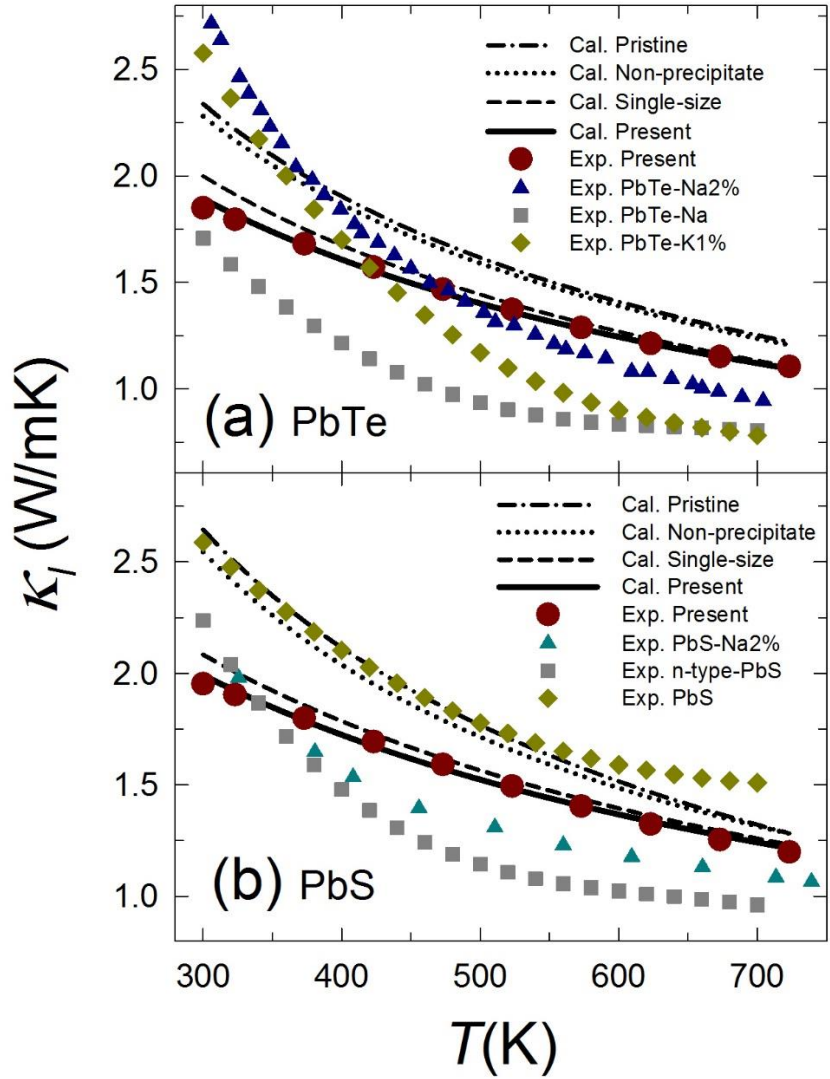

**Figure 4.** (a,b) Calculated lattice thermal conductivity (curves) *vs.* temperature in comparison with experimental data (symbols) for PbTe and PbS, respectively. The Cal. curve represents data computed using the model given in this work [Equation (8)]. The Cal. single-size curve represents data computed using the earlier single-size model [Equation (9)]. The Cal. non-precipitate curve represents data computed considering only the Umklapp, impurity and boundary scattering contributions to phonon scattering [Equation (2)]. The Cal. pristine curve represents data computed considering only the Umklapp and boundary scattering contributions.

We computed the values of the lattice thermal conductivity $\kappa_l$ of the PbTe and PbS samples explicitly taking the experimentally measured nanostructure size distributions (Figure 2f and grain size statistics provided in the Supporting Information) as input parameters for the above-described model [Equation (8)]. Figure 4 shows the temperature dependence of $\kappa_l$ in comparison with experimental measurements. We see that $\kappa_l$ is significantly reduced by nanoprecipitate phonon scattering, in particular for the PbS sample. This is probably because PbS has a larger mass density difference between the host and precipitate materials than PbTe, since the results of our high-magnification atomic imaging and energy dispersive spectroscopy analyses suggest that the precipitates are Pb-rich, as the mass density difference has long been considered as a major factor influencing the precipitate phonon scattering.[21] We remark that the experimentally-measured precipitate distributions lead to a decrease in $\kappa_l$ beyond the single-size limit calculated by the previously-proposed model. Moreover, we provide quantitative comparison with experimentally measured lattice thermal conductivity data in the literature[16, 37, 43-46], with values updated using Equations (11) - (14) for a valid comparison. Worthy of note is the fact that the experimentally measured thermal conductivities always include multiple contributions from different scattering sources, and it is hard to isolate the influence of precipitate scattering. It is in this context that theoretical approaches to the problem become the most valuable, especially in light of the lack of experimental microscopic information such as grain and precipitate size distributions.

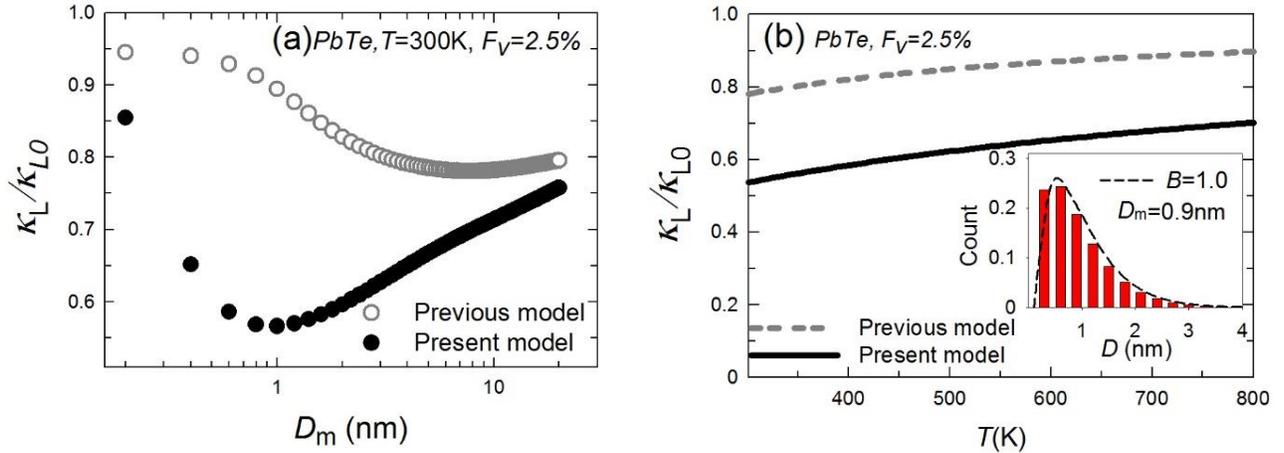

**Figure 5.** (a) Relative thermal conductivity vs. average size $D_m$ for a given distribution scale parameter $B = 1.0$ with a given the precipitate volumetric fraction $F_V$=2.5% at 300K. (b) Temperature-dependent relative thermal conductivities calculated by using the previous single-size model and the present model, respectively. The inset

shows the optimal precipitate size distribution leading to the lowest thermal conductivity. ($k_{l0}$ denotes the lattice thermal conductivity of pristine PbTe.)

It is important to investigate the effect of size distribution on the lattice thermal conductivity. The Gamma distribution is used here as a flexible approximation to the precipitate size distribution $f(D)$ [Equation (9)] for simplifying the theoretical analysis,

$$f(D) = \frac{D^{A-1}e^{-D/B}}{\Gamma(A)B^A}, \qquad (15)$$

where $A$ and $B$ are shape and scale parameters, respectively. $\Gamma$ is the usual generalized factorial

$$\Gamma(A) = \int_0^\infty t^{A-1}e^{-t}dt, \qquad (16)$$

The Gamma distribution is chosen because of its flexibility: both exponential and Gaussian distributions are comprised in this class, and it is also suitable for fitting distributions with "fat tails" (power-law decline). It is important to note that the mean diameter $D_m$ of the Gamma distribution can be written as

$$D_m = AB. \qquad (17)$$

Figure 5 shows the room-temperature lattice thermal conductivity of PbTe as a function of average size $D_m$ for a fixed scale parameter [Equation (13)]. We observe that the dependence of thermal conductivity on $D_m$ is similar to its dependence on particle size in the single-size approximation with an optimal size that minimizes $\kappa_l$ as shown in previous works.[24] However, the choice of $D_m$ does not exhaust the degrees of freedom in the problem. We thus observe that there exists an optimal size distribution that minimizes the thermal conductivity beyond the single-size limit. Furthermore, we perform a detailed exploration of the (A, B) space in search of an optimal size distribution that reduces the thermal conductivity most efficiently. Figure 5b shows the temperature dependence of the minimum relative thermal conductivities of the PbTe sample with Pb nanoprecipitates, calculated by considering an optimal size distribution within the limits of our parametric model. Pure Pb precipitates are used in the computation for simplicity since the results of Figure 3 suggest that the precipitates are Pb-rich while the exact composition of precipitates remains undetermined. It is readily seen that the thermal conductivity predicted under such an optimal size distribution is far lower than that given by the single-size approximation. This suggests that phonon scattering by nanoprecipitates in thermoelectric composites can be considerably stronger than previously suggested.

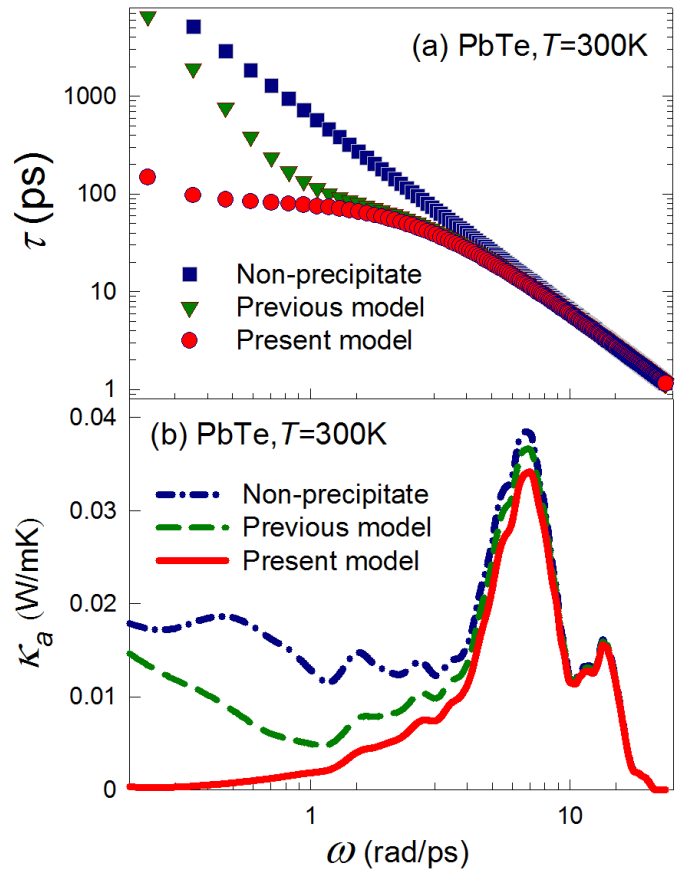

**Figure 6.** (a) Relaxation time as a function of phonon frequency. (b) Contribution to the lattice thermal conductivity from phonons of different frequency at 300 K.

To understand the effect of the size distribution on thermal conductivity, we compute the contribution to the total relaxation time as a function of the phonon frequency for different scattering cases (Figure 6a).

It can be seen that precipitate scattering mainly influences the low-frequency phonons from about 0.1 to 3.0 rad/ps. Precipitate scattering effectively reduces the thermal conductivity contribution of the low-frequency phonons (Figure 6b).

## CONCLUDING REMARKS

Bridging the gap between nanoscale phonon scattering and macroscopic thermoelectric transport properties requires accurate information about the nanostructure size distribution. Combining TEM experiments and theoretical phonon calculations we show that phonon scattering is dominated by nanoprecipitates for PbS samples with large grains, and thus that nanoprecipitates make an important contribution to bulk *ZT* enhancement by nanostructuring. Thermal conductivity is clearly found to depend on the precipitate size distribution and thus resists modeling in terms of a single size, as proposed in previous works. Moreover, the correlation between the distributed precipitate sizes and the phonons at different frequency/wavelength provides some physical basis for experimentally measured values of $\kappa_l$, and also a criterion to judge whether the thermal conductivity can be further optimized. These results can prove essential for *ZT* enhancement by nanostructuring in bulk thermoelectrics.

**ASSOCIATED CONTENT**

**Supporting Information**

The Supporting Information is available free of charge on the ACS Publications Web site. The Supporting Information is available free of charge on the ACS Publications website at DOI: xxxx/xxxx.

1. Calculation details, 2. Grain size measurements, 2. Grain size measurements, 4. Electron contribution to the thermal conductivity, 5. TEM images of nanoprecipitates.

**AUTHOR INFORMATION**

**Corresponding Authors**

*E-mail: he.jq@sustc.edu.cn (J.H.).

*E-mail: g.yang@mail.xjtu.edu.cn (G.Y.).

**ORCID**

Zhao Wang: 0000-0003-1887-223X

Jesus Carrete: 0000-0003-0971-1098

**Notes**

The authors declare no competing financial interest.

**ACKNOWLEDGMENTS**

This work is supported by the National Natural Science Foundation of China under Grant No. 51571007, the Guangxi Science Foundation (2013GXNSFFA019001), the Guangxi Key Laboratory Foundation (15-140-54), and the Scientific Research Foundation of GuangXi University (Grant No. XTZ160532).

# References

1. Snyder, G. J.; Toberer, E. S., Complex Thermoelectric Materials. *Nat Mater.* **2008,** *7*, 105-114.

2. Chen, Z.-G.; Han, G.; Yang, L.; Cheng, L.; Zou, J., Nanostructured Thermoelectric Materials: Current Research and Future Challenge. *Prog. Nat. Sci. : Mater. Int.* **2012,** *22*, 535-549.

3. Singh, A. K.; Raykar, V. S., Experimental Investigation of Thermal Conductivity and Effusivity of Ferrite Based Nanofluids under Magnetic Field. *ISRN Nanotech.* **2013**, 479763-479763.

4. Wu, H.; Carrete, J.; Zhang, Z.; Qu, Y.; Shen, X.; Wang, Z.; Zhao, L.-D.; He, J., Strong Enhancement of Phonon Scattering through Nanoscale Grains in Lead Sulfide Thermoelectrics. *NPG Asia Mater.* **2014,** *6*, e108.

5. He, J.; Zhao, L.-D.; Zheng, J.-C.; Doak, J. W.; Wu, H.; Wang, H.-Q.; Lee, Y.; Wolverton, C.; Kanatzidis, M. G.; Dravid, V. P., Role of Sodium Doping in Lead Chalcogenide Thermoelectrics. *J. Am. Chem. Soc.* **2013,** *135*, 4624-4627.

6. Korkosz, R. J.; Chasapis, T. C.; Lo, S.-h.; Doak, J. W.; Kim, Y. J.; Wu, C.-I.; Hatzikraniotis, E.; Hogan, T. P.; Seidman, D. N.; Wolverton, C.; Dravid, V. P.; Kanatzidis, M. G., High ZT in p-Type $(PbTe)_{1-2x}(PbSe)_x(PbS)_x$ Thermoelectric Materials. *J. Am. Chem. Soc.* **2014,** *136*, 3225-3237.

7. Zhao, L.-D.; Dravid, V. P.; Kanatzidis, M. G., The Panoscopic Approach to High Performance Thermoelectrics. *Energy Environ. Sci.* **2014,** *7*, 251-268.

8. Girard, S. N.; He, J.; Zhou, X.; Shoemaker, D.; Jaworski, C. M.; Uher, C.; Dravid, V. P.; Heremans, J. P.; Kanatzidis, M. G., High Performance Na-doped PbTe–PbS Thermoelectric Materials: Electronic Density of States Modification and Shape-Controlled Nanostructures. *J. Am. Chem. Soc.* **2011,** *133*, 16588-16597.

9. Minnich, A. J.; Dresselhaus, M. S.; Ren, Z. F.; Chen, G., Bulk Nanostructured Thermoelectric

Materials: Current Research and Future Prospects. *Energy Environ. Sci.* **2009,** *2*, 466-479.

10. Vineis, C. J.; Shakouri, A.; Majumdar, A.; Kanatzidis, M. G., Nanostructured Thermoelectrics: Big Efficiency Gains from Small Features. *Adv. Mater.* **2010,** *22*, 3970-3980.

11. Pei, Y.; Lensch-Falk, J.; Toberer, E. S.; Medlin, D. L.; Snyder, G. J., High Thermoelectric Performance in PbTe Due to Large Nanoscale $Ag_2Te$ Precipitates and La Doping. *Adv. Funct. Mater.* **2011,** *21*, 241-249.

12. Kanatzidis, M. G., Nanostructured Thermoelectrics: The New Paradigm? *Chem. Mater.* **2010,** *22*, 648-659.

13. Wang, H.; Pei, Y.; LaLonde, A. D.; Snyder, G. J., Heavily Doped p-Type PbSe with High Thermoelectric Performance: An Alternative for PbTe. *Adv. Mater.* **2011,** *23*, 1366-1370.

14. Sootsman, J. R.; Kong, H.; Uher, C.; D'Angelo, J. J.; Wu, C.-I.; Hogan, T. P.; Caillat, T.; Kanatzidis, M. G., Large Enhancements in the Thermoelectric Power Factor of Bulk PbTe at High Temperature by Synergistic Nanostructuring. *Angew. Chem. Int. Ed.* **2008,** *47*, 8618-8622.

15. Pei, Y.; LaLonde, A.; Iwanaga, S.; Snyder, G. J., High Thermoelectric Figure of Merit in Heavy Hole Dominated PbTe. *Energy Environ. Sci.* **2011,** *4*, 2085-2089.

16. Zhang, Q.; Cao, F.; Liu, W.; Lukas, K.; Yu, B.; Chen, S.; Opeil, C.; Broido, D.; Chen, G.; Ren, Z., Heavy Doping and Band Engineering by Potassium to Improve the Thermoelectric Figure of Merit in p-Type PbTe, PbSe, and $PbTe_{1-y}Se_y$. *J. Am. Chem. Soc.* **2012,** *134*, 10031-10038.

17. Kim, G. H.; Shao, L.; Zhang, K.; Pipe, K. P., Engineered Doping of Organic Semiconductors for Enhanced Thermoelectric Efficiency. *Nat. Mater.* **2013,** *12*, 719-723.

18. Caylor, J. C.; Coonley, K.; Stuart, J.; Colpitts, T.; Venkatasubramanian, R., Enhanced Thermoelectric Performance in PbTe-Based Superlattice Structures from Reduction of Lattice Thermal Conductivity. *Appl. Phys. Lett.* **2005,** *87*, 023105.

19. Pei, Y.; Heinz, N. A.; LaLonde, A.; Snyder, G. J., Combination of Large Nanostructures and

Complex Band Structure for High Performance Thermoelectric Lead Telluride. *Energy Environ. Sci.* **2011,** *4*, 3640-3645.

20. Zhang, Q.; Wang, H.; Liu, W.; Wang, H.; Yu, B.; Zhang, Q.; Tian, Z.; Ni, G.; Lee, S.; Esfarjani, K.; Chen, G.; Ren, Z., Enhancement of Thermoelectric Figure-of-Merit by Resonant States of Aluminium Doping in Lead Selenide. *Energy Environ. Sci.* **2012,** *5*, 5246-5251.

21. Yang, X.; Carrete, J.; Wang, Z., Role of Force-Constant Difference in Phonon Scattering by Nano-Precipitates in PbTe. *J. Appl. Phys.* **2015,** 118, 085701.

22. He, J.; Sootsman, J. R.; Girard, S. N.; Zheng, J.-C.; Wen, J.; Zhu, Y.; Kanatzidis, M. G.; Dravid, V. P., On the Origin of Increased Phonon Scattering in Nanostructured PbTe Based Thermoelectric Materials. *J. Am. Chem. Soc.* **2010,** *132*, 8669-8675.

23. Biswas, K.; He, J.; Zhang, Q.; Wang, G.; Uher, C.; Dravid, V. P.; Kanatzidis, M. G., Strained Endotaxial Nanostructures with High Thermoelectric Figure of Merit. *Nat. Chem.* **2011,** *3*, 160-166.

24. Mingo, N.; Hauser, D.; Kobayashi, N. P.; Plissonnier, M.; Shakouri, A., "Nanoparticle-in-Alloy" Approach to Efficient Thermoelectrics: Silicides in SiGe. *Nano Lett.* **2009,** *9*, 711-715.

25. Ahn, K.; Biswas, K.; He, J.; Chung, I.; Dravid, V.; Kanatzidis, M. G., Enhanced Thermoelectric Properties of p-Type Nanostructured PbTe-MTe (M = Cd, Hg) Materials. *Energy Environ. Sci.* **2013,** *6*, 1529-1537.

26. Biswas, K.; He, J.; Blum, I. D.; Wu, C.-I.; Hogan, T. P.; Seidman, D. N.; Dravid, V. P.; Kanatzidis, M. G., High-Performance Bulk Thermoelectrics with All-Scale Hierarchical Architectures. *Nature* **2012,** *489*, 414-418.

27. Ohta, M.; Biswas, K.; Lo, S.-H.; He, J.; Chung, D. Y.; Dravid, V. P.; Kanatzidis, M. G., Enhancement of Thermoelectric Figure of Merit by the Insertion of MgTe Nanostructures in p-Type PbTe Doped with $Na_2Te$. *Adv. Energy Mater.* **2012,** *2*, 1117-1123.

28. He, J.; Girard, S. N.; Kanatzidis, M. G.; Dravid, V. P., Microstructure-Lattice Thermal Conductivity

Correlation in Nanostructured PbTe$_{0.7}$S$_{0.3}$ Thermoelectric Materials. *Adv. Funct. Mater.* **2010,** *20*, 764-772.

29. Blachnik, R.; Igel, R., Thermodynamische Eigenschaften von IV−VI-Verbindungen: Bleichalkogenide/Thermodynamic Properties of IV−VI-Compounds: Leadchalcogenides. In *Z. Naturforsch. B*, **1974**, 29, 625.

30. Wang, Z.; Wang, S.; Obukhov, S.; Vast, N.; Sjakste, J.; Tyuterev, V.; Mingo, N., Thermoelectric Transport Properties of Silicon: Toward an *ab initio* Approach. *Phys. Rev. B* **2011,** *83*, 205208.

31. Wang, Z.; Mingo, N., Diameter Dependence of SiGe Nanowire Thermal Conductivity. *Appl. Phys. Lett.* **2010,** *97*, 101903.

32. Morelli, D. T.; Heremans, J. P.; Slack, G. A., Estimation of the Isotope Effect on the Lattice Thermal Conductivity of Group IV and Group III-V Semiconductors. *Phys. Rev. B* **2002,** *66*, 195304.

33. Cahill, D. G.; Braun, P. V.; Chen, G.; Clarke, D. R.; Fan, S.; Goodson, K. E.; Keblinski, P.; King, W. P.; Mahan, G. D.; Majumdar, A.; Maris, H. J.; Phillpot, S. R.; Pop, E.; Shi, L., Nanoscale Thermal Transport. II. 2003–2012. *Appl. Phys. Rev.* **2014,** 1, 011305.

34. Kim, W.; Majumdar, A., Phonon Scattering Cross Section of Polydispersed Spherical Nanoparticles. *J. Appl. Phys.* **2006,** *99*, 084306.

35. de Jong, M.; Chen, W.; Angsten, T.; Jain, A.; Notestine, R.; Gamst, A.; Sluiter, M.; Krishna Ande, C.; van der Zwaag, S.; Plata, J. J.; Toher, C.; Curtarolo, S.; Ceder, G.; Persson, K. A.; Asta, M., Charting the Complete Elastic Properties of Inorganic Crystalline Compounds. *Sci. Data* **2015,** *2*, 150009.

36. Zhang, H.; Minnich, A. J., The Best Nanoparticle Size Distribution for Minimum Thermal Conductivity. *Sci. Rep.* **2015,** *5*, 8995.

37. Pei, Y.-L.; Liu, Y., Electrical and Thermal Transport Properties of Pb-Based Chalcogenides: PbTe, PbSe, and PbS. *J. Alloys Compd.* **2012,** *514*, 40-44.

38. Fang, H.; Luo, Z.; Yang, H.; Wu, Y., The Effects of the Size and the Doping Concentration on the

Power Factor of n-Type Lead Telluride Nanocrystals for Thermoelectric Energy Conversion. *Nano Lett.* **2014,** *14*, 1153-7.

39. May, A. F.; Toberer, E. S.; Saramat, A.; Snyder, G. J., Characterization and Analysis of Thermoelectric Transport in n-Type $Ba_8Ga_{16-x}Ge_{30+x}$. *Phys. Rev. B* **2009,** 80, 125205.

40. Zhang, Q.; Cao, F.; Lukas, K.; Liu, W.; Esfarjani, K.; Opeil, C.; Broido, D.; Parker, D.; Singh, D. J.; Chen, G.; Ren, Z., Study of the Thermoelectric Properties of Lead Selenide Doped with Boron, Gallium, Indium, or Thallium. *J. Am. Chem. Soc.* **2012,** *134*, 17731-17738.

41. Wang, H.; Gibbs, Z. M.; Takagiwa, Y.; Snyder, G. J., Tuning Bands of PbSe for Better Thermoelectric Efficiency. *Energy Environ. Sci.* **2014,** *7*, 804-811.

42. LaLonde, A. D.; Pei, Y.; Snyder, G. J., Reevaluation of $PbTe_{1-x}I_x$ as High Performance n-Type Thermoelectric Material. *Energy Environ. Sci.* **2011,** *4*, 2090-2096.

43. Wang, H.; Schechtel, E.; Pei, Y.; Snyder, G. J., High Thermoelectric Efficiency of n-Type PbS. *Adv. Energy Mater.* **2013,** *3*, 488-495.

44. Pei, Y.; Gibbs, Z. M.; Gloskovskii, A.; Balke, B.; Zeier, W. G.; Snyder, G. J., Optimum Carrier Concentration in n-Type PbTe Thermoelectrics. *Adv. Energy Mater.* **2014,** 4, 1400486.

45. Lo, S.-H.; He, J.; Biswas, K.; Kanatzidis, M. G.; Dravid, V. P., Phonon Scattering and Thermal Conductivity in p-Type Nanostructured PbTe-BaTe Bulk Thermoelectric Materials. *Adv. Funct. Mater.* **2012,** *22*, 5175-5184.

46. Pei, Y.; May, A. F.; Snyder, G. J., Self-Tuning the Carrier Concentration of $PbTe/Ag_2Te$ Composites with Excess Ag for High Thermoelectric Performance. *Adv. Funct. Mater.* **2011,** 1, 291-296.

# Table of Contents Graphic

**ABSTRACT:** We present a comprehensive experimental and theoretical study of phonon scattering by nanoprecipitates in potassium-doped PbTe, PbSe and PbS. We highlight the role of the precipitate size distribution measured by microscopy, whose tuning allows for thermal conductivities lower than the limit achievable with a single size. The correlation between the size distribution and the contributions to thermal conductivity from phonons in different frequency ranges provides a physical basis to the experimentally measured thermal conductivities, and a criterion to estimate the lowest achievable thermal conductivity. The results have clear implications for efficiency enhancements in nanostructured bulk thermoelectrics.

**KEYWORDS:** thermoelectric; precipitate interface; phonon; thermal transport; lead chalcogenide.

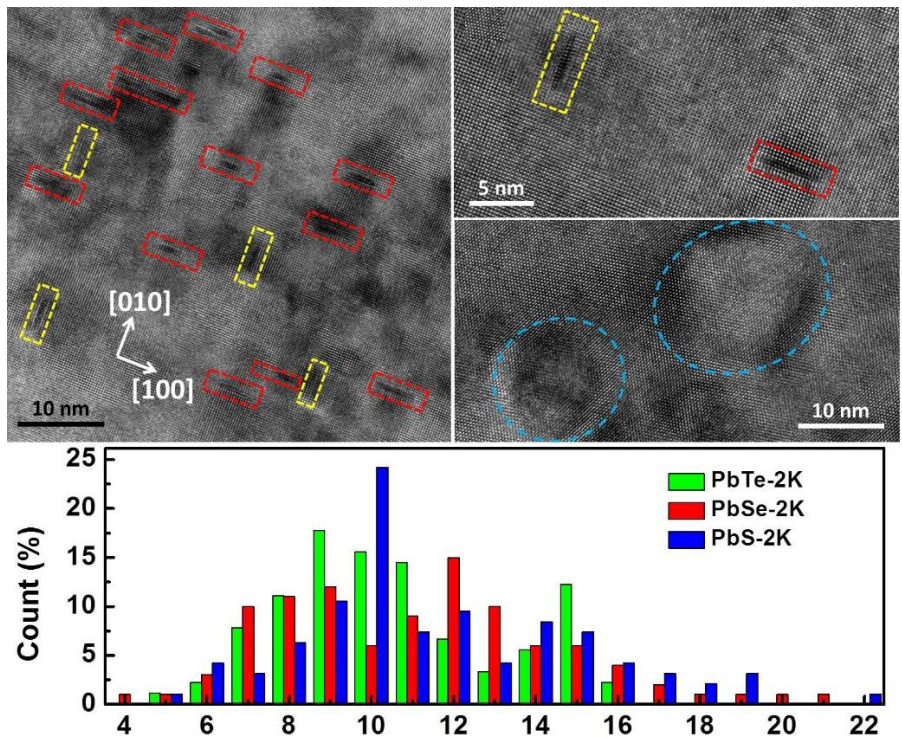